\documentclass[preprintnumbers,superscriptaddress,pra]{revtex4}%
\usepackage[colorlinks=true,linkcolor=blue]{hyperref}
\usepackage{amssymb}
\usepackage{amsfonts}
\usepackage{amsmath}
\usepackage{graphicx}%
\providecommand{\U}[1]{\protect\rule{.1in}{.1in}}

\let\stdsection\section
\renewcommand\section{\nopagebreak\stdsection}
\begin{document}
\title{Gravitation induced shrinkage of Mercury's orbit }
\author{Qian Moxian}
\affiliation{College of physics and electronic information, Inner Mongolia Normal
University, 010022 Hohhot, China}
\author{Li Xibin}
\affiliation{College of physics and electronic information, Inner Mongolia Normal
University, 010022 Hohhot, China}
\author{Cao Yongjun}
\email{phyjcao@imnu.edu.cn}
\affiliation{College of physics and electronic information, Inner Mongolia Normal
University, 010022 Hohhot, China}

\begin{abstract}
In general relativity, the Mercury's orbit becomes approximately elliptical
and the its perihelion has thus an additional advance. We demonstrate,
meanwhile, that in comparison of those given by the Newton's theory of
gravitation for the orbit of the Mercury, the circumference and the area are
reduced by $40.39$ km and $2.35\ast10^{9}$ km$^{2}$, respectively, beside the
major-axis contraction pointed out recently, and all are produced by the
curved space within the Einstein's theory of gravitation. Since the resolution
power of present astronomical distance measurement technology reaches $one$
kilometer, the shrinkage of Mercury's orbit can then be observable.
\end{abstract}
\maketitle

\section{Introduction}

\label{sec:intro}

The gravitational field produced by the Sun causes the space nearby to be
appreciably curved. Since the Mercury is the planet closest to the Sun, the
general relativity provides a more accurate description of its orbit than what
the Newton's law of universal gravitation does. The Mercury's perihelion
advance, $43$ arcseconds per century added to the planet's orbital precession,
was very earliest successful prediction of the general relativity.
\cite{a,b,c} In fact, the advance has long been one of the observational
cornerstones of general relativity, under more and more stringent theoretical
examinations and experimental tests to date. \cite{d1,d2} The gravitational
interaction between the Mercury and the Sun does not only result in the
perihelion advance of the orbit of Mercury, but also the radial contraction.
\cite{l} Recently, two new potentially observable effects on the orbit of
Mercury are reported. \cite{d2,l} One is related to the angular coordinate,
the second order effect of the perihelion advance, which adds $1$ degree of
arc every two billion years, believed to be detectable by BepiColombo, a
European and Japanese mission to Mercury launched on 20 October 2018.
\cite{d2} Another is related to the radial coordinate, the major-axis
contraction of Mercury's elliptical orbit by an amount of $1.3$ km, which is
within the limit of resolution power of current astronomical distance
measurement technology which can be seen from the website of NASA where the
distance of the Mercury from the Sun is updated every second with accuracy of
$one$ kilometer. \cite{f,g} We are confident that the major-axis contraction
is one aftereffect of an overall shrinkage of the Mercury's orbit given by the
general relativity in contrast to that given by the Newton's theory of
gravitation. By an overall shrinkage of an ellipse, we mean that both the
circumference and the area are reduced to some extents, though in general
relativity the Mercury's orbit is approximately elliptical.

In Sec.~\ref{sec2}, the basics of the Mercury's orbit are outlined. In
Sec.~\ref{sec3}, the circumference of the Mercury's orbit is explicitly
calculated and illustrated to be shorter than that given by the Newton's
theory. In Sec.~\ref{sec4}, the area of the Mercury's orbit is defined and
then shown to be less than that given by the Newton's theory. The last section
\ref{sec5} is a brief conclusion.

\section{Basics of the Mercury's orbit}

\label{sec2}

In the Newton's law of universal gravitation, the space-time is flat, and the
trajectory of a planet circling the Sun follows a perfect equation of the
ellipse,
\begin{equation}
R(\theta)=\frac{r_{\text{gs}}}{b^{2}}\frac{1}{1+e\cos\theta}, \label{ellipse}%
\end{equation}
where $R$ is the radial coordinate in the plane polar coordinate
system,~$\theta$~is the angular coordinate, $e$ is centricity,~$b=GM/cL$~is a
dimensionless characteristic quantity, $L$ is the angular momentum of a unit
mass planet, $c$ is the speed of light, and~$r_{\text{gs}}=GM/c^{2}%
\approx1.47$~km is the Schwarzschild radius of the Sun. For the Mercury,
$e=0.206$, and $b=1.63\ast10^{-4}$. \cite{f,g}

In the Einstein's gravitational theory, the space-time is curved, and the
space in the solar system is described by the Schwarzschild metric with the
Sun at the origin $r=0$,%
\begin{equation}
ds^{2}=-\left(  1-\frac{2r_{\text{gs}}}{r}\right)  dt^{2}+\frac{dr^{2}%
}{1-2r_{\text{gs}}/r}+r^{2}\left(  d\theta^{2}+\sin^{2}\theta d\phi
^{2}\right)  . \label{metric}%
\end{equation}
where $\left(  r,\theta,\phi\right)  $ are spherical polar coordinates, and
$ds$ is the infinitesimal proper length element. It is evidently that once
$r\gg r_{\text{gs}}$, the space is asymptotically flat and only the orbit of
the Mercury, the innermost planet in the solar system, is noticeably distorted
from (\ref{ellipse}) as, \cite{l}
\begin{equation}
r\approx r_{\text{gs}}\left[  \frac{1}{b^{2}(1+e\cos t)}-\frac{2(3+2e^{2}%
+e^{2}\cos t)}{(1+e\cos t)^{2}}\sin^{2}(\frac{t}{2})\right]
=R(t)-r_{\text{gs}}f(t), \label{radial}%
\end{equation}
where
\begin{equation}
f(t)=\frac{2(3+2e^{2}+e^{2}\cos t)}{(1+e\cos t)^{2}}\sin^{2}\left(  \frac
{t}{2}\right)  \label{f}%
\end{equation}
and $t$ is a parameter related to angular variable $\theta$ via,
\begin{equation}
t(\theta)\approx(1-3b^{2})\theta,\text{or},\theta\approx(1+3b^{2})t(\theta).
\label{radial1}%
\end{equation}
The amount of the procession of Mercury's perihelion in a period $\Delta
t=2\pi$ is therefore,
\begin{equation}
\Delta\theta\approx3b^{2}\Delta t(\theta)=6\pi b^{2}.
\end{equation}
It is the classic result \cite{a,b,c} which in a century of the Earth gives
$43$ arcseconds. The second order angular procession is of order
$O(b^{4})=O(10^{-15})$. \cite{d2} From relation (\ref{radial1}), it is worth
stressing that $t\in\lbrack0,2\pi]$ is equivalent to $\theta\in\lbrack
0,2\pi+6\pi b^{2}]$.

The infinitesimal line element $d\ell$ along the trajectory of the Mercury is
in flat space
\begin{equation}
d\ell=\sqrt{dx^{2}+dy^{2}}=\sqrt{dR^{2}+(Rd\theta)^{2}}=\sqrt{R^{2}+\left(
\frac{dR}{d\theta}\right)  ^{2}}d\theta.
\end{equation}
With $R$ given by (\ref{ellipse}) the circumference of Mercury's orbit in
Newton's theory of gravitation is,
\begin{equation}
\ell_{n}=\int_{0}^{2\pi}\sqrt{R^{2}+\left(  \frac{dR}{d\theta}\right)  ^{2}%
}d\theta. \label{newtonC}%
\end{equation}
Similarly, the area $S_{n}$ of the ellipse is,%
\begin{equation}
S_{n}=\frac{1}{2}\int_{0}^{2\pi}R(\theta)d\ell=\frac{1}{2}\int_{0}^{2\pi
}R(\theta)\sqrt{\left(  \frac{dR}{d\theta}\right)  ^{2}+R^{2}}d\theta.
\label{newtonS}%
\end{equation}
In Einstein's theory of gravitation, the circumference $\ell$ and the area $S$
must be defined by, respectively, the length of the trajectory of the Mercury
and the area swept by a moving line connected the Mercury and the Sun, from
one perihelion to the next one, from which the angular parameter $\theta$
changes from $0$ to $2\pi+6\pi b^{2}$, i.e., $t$ changes from $0$ to $2\pi$
from (\ref{radial1}), Two differences $\ell-\ell_{n}$ and $S-S_{n}$ define,
respectively, the changes of the circumference and the area of Mercury's orbit.

\section{Calculation of the circumference difference $\ell-\ell_{n}$}

\label{sec3}

The space-time of around the Sun is static, and the spatially separated clocks
are synchronized in metric (\ref{metric}). The infinitesimal proper distance
element~$d\ell$ along the circumference of the Mercury's orbit is from
(\ref{metric})
\begin{equation}
d\ell^{2}=\frac{1}{1-2r_{\text{gs}}/r}dr^{2}+r^{2}d\theta^{2},
\end{equation}
i.e.,
\begin{equation}
d\ell=\sqrt{\frac{1}{1-2r_{\text{gs}}/r}\left(  \frac{dr}{d\theta}\right)
^{2}+r^{2}}\ d\theta, \label{perimeter}%
\end{equation}
and
\begin{equation}
\ell=\int_{0}^{2\pi+6\pi b^{2}}\sqrt{\frac{1}{1-2r_{\text{gs}}/r}\left(
\frac{dr}{d\theta}\right)  ^{2}+r^{2}}\ d\theta=\int_{0}^{2\pi}\sqrt{\frac
{1}{1-2r_{\text{gs}}/r}\left(  \frac{dr}{dt}\right)  ^{2}+r^{2}}\ dt,
\label{Cir3}%
\end{equation}
To determine the length to the order of $\mathcal{O}(r_{\text{gs}})$, we need
to expand $r^{2}$ and $\left(  dr/dt\right)  ^{2}$ to this order. We have from
(\ref{radial}) and (\ref{radial1}),
\begin{subequations}
\begin{align}
r^{2}  &  \approx R(t)^{2}-2r_{\text{gs}}R(t)f(t),\text{and}
\label{perimeter1}\\
\frac{dr}{dt}  &  \approx\frac{dR(t)}{dt}-r_{\text{gs}}\frac{df(t)}%
{dt},\text{and}\label{perimeter1-1}\\
\left(  \frac{dr}{dt}\right)  ^{2}  &  \approx\left(  \frac{dR(t)}{dt}\right)
^{2}-2r_{\text{gs}}\frac{dR(t)}{dt}\frac{df(t)}{dt}. \label{perimeter1-2}%
\end{align}
The approximate expression for $\left(  1-2r_{\text{gs}}/r\right)
^{-1}\left(  dr/dt\right)  ^{2}$ to the order of $\mathcal{O}(r_{\text{gs}})$
is
\end{subequations}
\begin{equation}
\frac{1}{1-2r_{\text{gs}}/r}\left(  \frac{dr}{dt}\right)  ^{2}\approx\left(
\frac{dR(t)}{dt}\right)  ^{2}-2r_{\text{gs}}\frac{dR(t)}{dt}\frac{df(t)}%
{dt}+2r_{\text{gs}}\frac{1}{R(t)}\left(  \frac{dR(t)}{dt}\right)  ^{2}.
\label{perimeter2}%
\end{equation}
Combining two results (\ref{perimeter1}) and (\ref{perimeter2}), we have
\begin{equation}
\frac{1}{1-2r_{\text{gs}}/r}\left(  \frac{dr}{dt}\right)  ^{2}+r^{2}%
\approx\left(  \frac{dR(t)}{dt}\right)  ^{2}+R(t)^{2}-2r_{\text{gs}}Q(t),
\end{equation}
where%
\begin{equation}
Q(t)=\frac{dR(t)}{dt}\frac{df(t)}{dt}+R(t)f(t)-\frac{1}{R(t)}\left(
\frac{dR(t)}{dt}\right)  ^{2}.
\end{equation}
We obtain%
\begin{equation}
\sqrt{\frac{1}{1-2r_{\text{gs}}/r}\left(  \frac{dr}{dt}\right)  ^{2}+r^{2}%
}\approx\sqrt{\left(  \frac{dR}{dt}\right)  ^{2}+R^{2}}-r_{\text{gs}}%
\cdot\dfrac{Q(t)}{\sqrt{\left(  \frac{dR}{dt}\right)  ^{2}+R^{2}}}.
\label{perimeter3}%
\end{equation}

Now we are in position to compute $\ell$ (\ref{Cir3}) which is given by%
\begin{align}
\ell &  \approx\int_{0}^{2\pi}\sqrt{\frac{1}{1-2r_{\text{gs}}/r}\left(
\frac{dr}{dt}\right)  ^{2}+r^{2}}dt\nonumber\\
&  \approx\int_{0}^{2\pi}\sqrt{\left(  \frac{dR}{dt}\right)  ^{2}+R^{2}%
}dt-r_{\text{gs}}\int_{0}^{2\pi}\dfrac{Q(t)}{\sqrt{\left(  \frac{dR}%
{dt}\right)  ^{2}+R^{2}}}dt\nonumber\\
&  \approx\ell_{n}-r_{\text{gs}}\int_{0}^{2\pi}\dfrac{Q(t)}{\sqrt{\left(
\frac{dR}{dt}\right)  ^{2}+R^{2}}}dt.
\end{align}
To get the difference $\Delta\ell\equiv\ell-\ell_{n}$ from above equation, we
need to compute the last integration in it, which can be easily carried out
numerically,
\begin{equation}
\int_{0}^{2\pi}\dfrac{Q(t)}{\sqrt{\left(  \frac{dR}{dt}\right)  ^{2}+R^{2}}%
}dt\approx27.48.
\end{equation}
We find immediately
\begin{equation}
\Delta\ell=-r_{\text{gs}}\int_{0}^{2\pi}\dfrac{Q(t)}{\sqrt{\left(  \frac
{dR}{dt}\right)  ^{2}+R^{2}}}dt\approx-27.48r_{gs}\approx-40.39\ \text{km.}
\label{deltal}%
\end{equation}
Thus, the space curvature makes the circumference of Mercury's orbit shortened.

\section{Calculation of the area difference $S-S_{n}$}

\label{sec4}

During one period $\theta\in\left[  0,2\pi+6\pi b^{2}\right]  $, a line drawn
from Mercury to the Sun sweeps out an area as Mercury moves. The infinitesimal
area element is defined by,%
\begin{equation}
dS=\frac{1}{2}rd\ell.
\end{equation}
The formula of calculation of the area is then from (\ref{radial1}),%
\begin{equation}
S=\frac{1}{2}\int_{0}^{2\pi+6\pi b^{2}}rd\ell(\theta)=\frac{1}{2}\int%
_{0}^{2\pi}rd\ell(t). \label{S}%
\end{equation}
In evaluation of this integral for the Mercury, we need to expand
$rd\ell(t)/dt$ in it to order of $\mathcal{O}(r_{\text{gs}})$ from
(\ref{Cir3})
\begin{align}
r\frac{d\ell(t)}{dt}  &  =r\sqrt{\frac{1}{1-2r_{\text{gs}}/r}\left(  \frac
{dr}{dt}\right)  ^{2}+r^{2}}\nonumber\\
&  \approx\left(  R(t)-r_{\text{gs}}f(t)\right)  \left(  \sqrt{\left(
\frac{dR}{dt}\right)  ^{2}+R^{2}}-r_{\text{gs}}\cdot\dfrac{Q(t)}{\sqrt{\left(
\frac{dR}{dt}\right)  ^{2}+R^{2}}}\right) \nonumber\\
&  =R(t)\sqrt{\left(  \frac{dR}{dt}\right)  ^{2}+R^{2}}-r_{\text{gs}}\left(
\dfrac{R(t)Q(t)}{\sqrt{\left(  \frac{dR}{dt}\right)  ^{2}+R^{2}}}%
+f(t)\sqrt{\left(  \frac{dR}{dt}\right)  ^{2}+R^{2}}\right)  .
\end{align}
The expression of the area $S$ to the order $\mathcal{O}(r_{\text{gs}})$ is
then
\begin{align}
S  &  \approx\frac{1}{2}\int_{0}^{2\pi}R(t)\sqrt{\left(  \frac{dR}{dt}\right)
^{2}+R^{2}}dt-r_{\text{gs}}\frac{1}{2}\int_{0}^{2\pi}\left(  \dfrac
{R(t)Q(t)}{\sqrt{\left(  \frac{dR}{dt}\right)  ^{2}+R^{2}}}+f(t)\sqrt{\left(
\frac{dR}{dt}\right)  ^{2}+R^{2}}\right)  dt\nonumber\\
&  \approx S_{n}-r_{\text{gs}}G, \label{S-Sn}%
\end{align}
where $r_{\text{gs}}G$ is the first order correction from Einstein's theory
with $G$ denoting an integral,
\begin{equation}
G=\frac{1}{2}\int_{0}^{2\pi}\left(  \dfrac{R(t)Q(t)}{\sqrt{\left(  \frac
{dR}{dt}\right)  ^{2}+R^{2}}}+f(t)\sqrt{\left(  \frac{dR}{dt}\right)
^{2}+R^{2}}\right)  dt,
\end{equation}
which can be numerically determined to be%
\begin{equation}
G\approx28.78\frac{r_{\text{gs}}}{b^{2}}.
\end{equation}
In final, we get the amount of the area reduction, $\Delta S=S-S_{n}%
\approx-r_{\text{gs}}G$, with $b=$ $1.63\ast10^{-4}$ and $r_{\text{gs}}%
=1.47$~km, which is simply,
\begin{equation}
\Delta S=-28.78\frac{r_{\text{gs}}^{2}}{b^{2}}=-2.35\ast10^{9}\text{ km}^{2}.
\label{deltas}%
\end{equation}
It is nearly the same area of a rectangle formed by one longer side length
which is $57.91\ast10^{6}$ km which is the semimajor axis of the Mercury
\cite{f,g}, and one shorter side length which is $\left\vert \Delta
\ell\right\vert =$ $40.39\ $km (\ref{deltal}). Result (\ref{deltas}) offers
another potentially observable gravitational effect induced by the
gravitational interaction between the Sun and the Mercury.

\section{Conclusion}

\label{sec5}

It is an important issue to search for a new possible experimental hunt of a
general relativistic effect. We report that the gravitational interaction
between the Sun and the Mercury produces an overall shrinkage of the Mercury's
orbit in comparison with that predicted by the Newton's theory of the
gravitation. Explicitly, the circumference $C$ is reduced by an amount
$40.39\ $km, and the covering area $S$ of the orbit in a period is diminished
by $2.35\ast10^{9}\ $km$^{2}$. With development of the modern technologies of
measuring the sizes in astronomical scale, such a shrinkage seems to fall in
the resolution power of the current apparatus.

\end{document}